\documentclass[nofootinbib,superscriptaddress,a4paper,twocolumn]{revtex4-1}

\usepackage{geometry}
\usepackage[english]{babel}
\usepackage[latin1]{inputenc}
\usepackage{hyperref}
\usepackage{amsmath, amssymb, amsfonts, mathrsfs}
\usepackage{graphicx}
\usepackage{xspace}
\newcommand{\ket}[1]{|#1\rangle}

\newcommand*{\melvin}{{\small M}{\scriptsize ELVIN}\xspace}

\begin{document} 

\title{Twisted Photons: New Quantum Perspectives in High Dimensions}
\author{Manuel Erhard}
\email{manuel.erhard@univie.ac.at}
\affiliation{Vienna Center for Quantum Science \& Technology (VCQ), Faculty of Physics, University of Vienna, Boltzmanngasse 5, 1090 Vienna, Austria.}
\affiliation{Institute for Quantum Optics and Quantum Information (IQOQI), Austrian Academy of Sciences, Boltzmanngasse 3, 1090 Vienna, Austria.}
\author{Robert Fickler}
\email{rfickler@uottawa.ca}
\affiliation{Department of Physics, University of Ottawa, Ottawa, ON, K1N 6N5, Canada.}
\author{Mario Krenn}
\email{mario.krenn@univie.ac.at}
\affiliation{Vienna Center for Quantum Science \& Technology (VCQ), Faculty of Physics, University of Vienna, Boltzmanngasse 5, 1090 Vienna, Austria.}
\affiliation{Institute for Quantum Optics and Quantum Information (IQOQI), Austrian Academy of Sciences, Boltzmanngasse 3, 1090 Vienna, Austria.}
\author{Anton Zeilinger}
\email{anton.zeilinger@univie.ac.at}
\affiliation{Vienna Center for Quantum Science \& Technology (VCQ), Faculty of Physics, University of Vienna, Boltzmanngasse 5, 1090 Vienna, Austria.}
\affiliation{Institute for Quantum Optics and Quantum Information (IQOQI), Austrian Academy of Sciences, Boltzmanngasse 3, 1090 Vienna, Austria.}
\maketitle 
\tableofcontents
\section{General Introduction}
\begin{figure*}[htbp]
	\centering
	\includegraphics[width=0.95\textwidth]{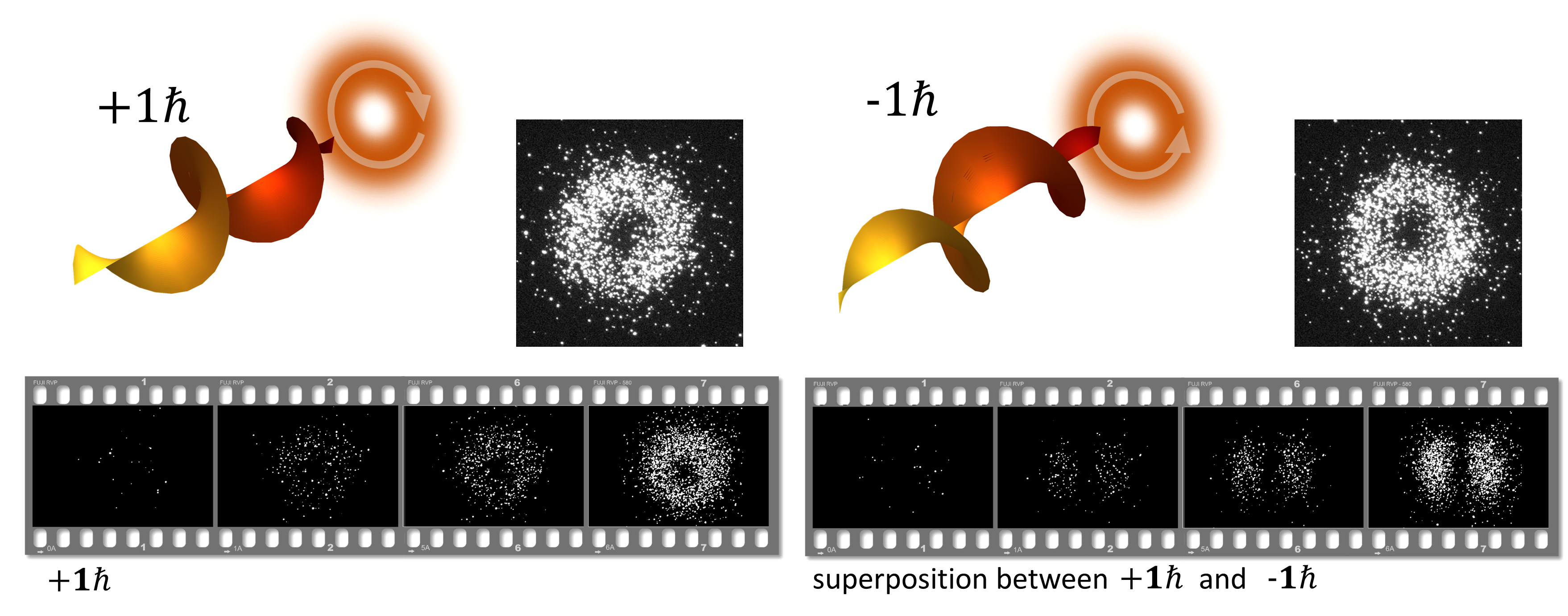}
	\caption{Orbital angular momentum (OAM) of twisted photons. A helical (twisted) phase $e^{il\phi}$ structure leads to a quantized amount of $l\hbar$ of OAM. Upper row: Photons with $+1\hbar$ of OAM have a phase structure that varies azimuthally from $0$ to $2\pi$, which also leads to a vortex along the beam axis and thus an intensity null. The black/white images are build by accumulating many single photon events. Lower row: Accumulation of approximately 30, 250, 600 and 2500 photons (from left to right) to reveal the intensity structure of an $\ket{+1}$ OAM mode and the superposition between $+1\hbar$ and $-1\hbar$. The images have been recorded with a single-photon sensitive, low-noise ICCD camera \cite{fickler2013real}.}
\label{fig:lg-modes}
\end{figure*}
Quantum information science and quantum information technology have seen a virtual explosion world-wide. It is all based on the observation that fundamental quantum phenomena on the individual particle or system-level lead to completely novel ways of encoding, processing and transmitting information. Quantum mechanics, a child of the first third of the 20th century, has found numerous realizations and technical applications, much more than was thought at the beginning. Decades later, it became possible to do experiments with individual quantum particles and quantum systems. This was due to technological progress, and for light in particular, the development of the laser. Only about 40 years ago, quantum phenomena were studied for their own curiosity. The motivation was that the predictions of quantum mechanics for individual systems were rather counterintuitive. People wanted to see these in a concrete way in the laboratory. Fundamental phenomena include quantum superposition for individual particles, as corroborated for example by neutron interferometers and by the double slit experiment for individual massive particles and for individual photons, the objective randomness of the individual quantum event, quantum entanglement \cite{schrodinger1935gegenwartige} and entanglement as signified by the Einstein-Podolsky-Rosen paradox \cite{einstein1935can} and Bell's theorem \cite{bell1964einstein} or the no-cloning theorem \cite{wootters1982single}. To date, these fundamental experiments in the new field of quantum information science and quantum information technology were signified by such notions as quantum communication, quantum cryptography, quantum teleportation, quantum computation and many more. It is a broad understanding that all this will lead to a future new information technology.

Hitherto, nearly all experiments and also nearly all realizations in the fields just mentioned have been performed with qubits, which are two-level quantum systems. We suggest that this limitation is again mainly a technological one, because it is very difficult to create, manipulate and measure more complex quantum systems.

Here, we provide a specific overview of some recent developments with higher-dimensional quantum systems. We mainly focus on Orbital Angular Momentum (OAM) states of photons and possible applications in quantum information protocols. Such states form discrete higher-dimensional quantum systems, also called qudits. A qudit can be seen as a quantum particle which is not limited to two states but in principle can have any number of discrete levels.

Specifically, we will first address the question what kind of new fundamental properties exist and the quantum information applications which are opened up by such novel systems. Then we give an overview of recent developments in the field by discussing several notable experiments over the past 2-3 years. Finally, we conclude with several important open questions which will be interesting for investigations in the future.

Due to limitation of space, we cannot cover breakthrough experiments in the area of high-dimensional quantum information memories~\cite{ding2016high,parigi2015storage}, in high-dimensional quantum information processing with other physical systems~\cite{kiktenko2015multilevel,kiktenko2015single,schaeff2015experimental,martin2017quantifying}, coherent transitions of atomic states driven by the OAM of photons \cite{schmiegelow2016transfer} or novel methods to manipulate OAM of light which might improve future quantum experiments, such as plasmonic q-plates \cite{karimi2014generating} and light carrying OAM produced in waveguide chips \cite{cai2012integrated}.

As we focus here on the very recent developments in this field, we have to leave out many important historic experiments. Readers interested in the early developments in quantum science with OAM of photons, we would like to refer to a review from 2007 \cite{molina2007twisted} and from 2017 \cite{krenn2017orbital}. Readers interested in the theoretical foundations of quantum information may look into \cite{nielsen2010quantum,horodecki2009quantum,plenio2005introduction,bruss2002characterizing}, and to get an overview of photonic quantum information one may look into \cite{multi-photon-review-zeilinger}.

\section{OAM of photons}

\begin{figure*}[htbp]
	\centering
	\includegraphics[width=0.95\textwidth]{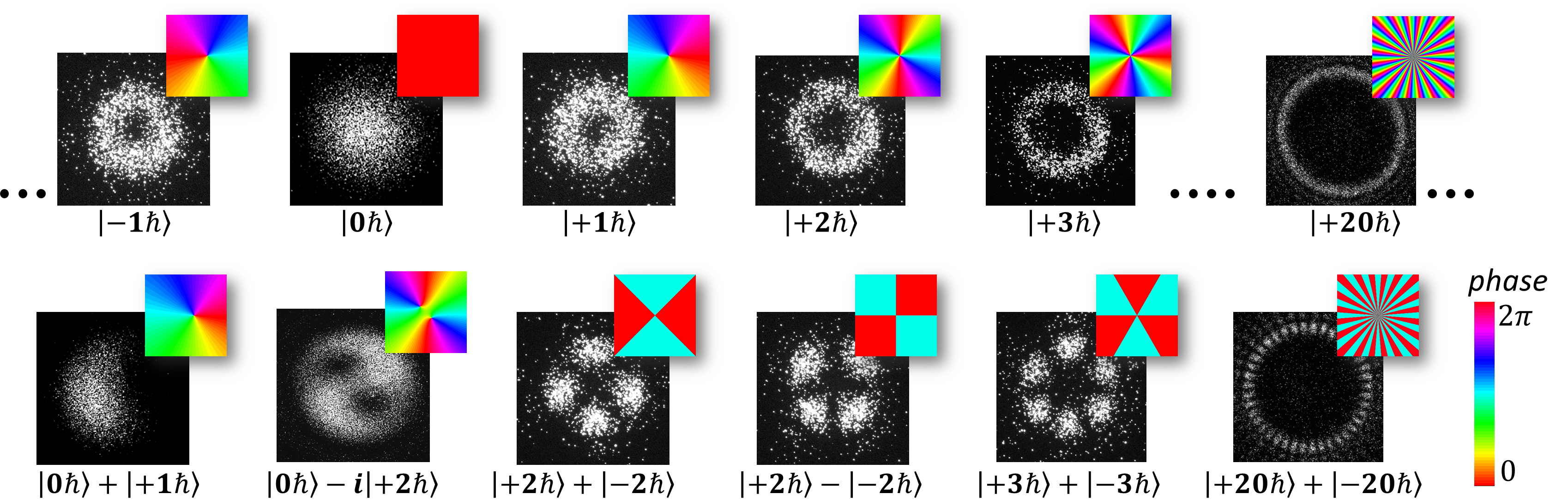}
	\caption{Single photon images of higher order OAM states (upper row) and superpositions states (lower row). Each mode can be used to increase dimensionality of the qudit state, e.g. the subset modes in the upper row from $-1\hbar$ to $=3\hbar$ could describe a five dimensional state. Quantum experiments also require coherent superposition states, a few examples of specific modes are shown in the lower row. All insets display the underlying phase structure in the color scale shown on the lower right.}
\label{fig:QuditSuperpos}
\end{figure*}

Photons are excellent for quantum communication as they can be transmitted over long distances with very low loss and without any known decoherence mechanism in free space \cite{yin2017satellite}. A special property of photons is their orbital angular momentum~\cite{allen1992orbital}. Quantum mechanically, the orbital angular momentum occurs in discrete steps of $\ell\hbar$, where $\ell$ is an im principle unbounded integer. Therefore, high-dimensional quantum information can be stored in the orbital angular momentum of single photons.

The electromagnetic field of a Laguerre-Gauss mode $\text{LG}^\ell_p$ accurately describes a photon carrying an OAM of $\ell\hbar$. Figure~\ref{fig:lg-modes} shows experimental images of single photons with different OAM values. The defining feature of LG modes is their spatial phase pattern. In particular, it is a helical phase that wraps around the axis of propagation $\ell$ times, which results in a phase singularity at the beam center. Thus the intensity profile of an LG mode shows the typical donut shaped structure. For single photons, the intensity profile gives the probability to detect a photon at a certain point. If we place a triggered single-photon camera and detect many heralded photons \cite{fickler2013real}, then the intensity profile of the respective LG mode emerges, which is shown in the filmstrips in Fig.~\ref{fig:lg-modes}. A few examples of higher order OAM modes, which can be used to encode high-dimensional qudit states, together with some exemplary superpositions structures are shown in Figure \ref{fig:QuditSuperpos}. 

\section{Advantages of Higher Dimensional Quantum Systems}

\subsection{Higher Information Capacity}
The first and rather obvious benefit of qudits is the enlarged information content a single quantum carrier can transmit. Any additional orthogonal state used in the encoding scheme enlarges the information encoded in a single quantum system. A simple example would be the use of ququarts, i.e. four dimensional systems, which can be used to encode 2 bits of information: $\ket{0}= 00$, $\ket{1}= 01$, $\ket{2}= 10$ and $\ket{3}= 11$. The often used measure of information capacity, which is given in $\log_2(d)$ bits/photon, can be seen as a measure of how many classical bits have to be used to be able to encode the same amount of information. In principle there is no upper limit of $d$ \footnote{Except of a potential limit which arises due to the finite size of the universe.}, which means that a single quantum system, like a photon, has the potential to encode an arbitrary large amount of information. 

\subsection{Enhanced robustness against eavesdropping and quantum cloning} \label{chapterSecurityQC}
Although perfect copying of a quantum state is impossible due to the no-cloning theorem, it was discovered that it is possible to approximatively clone a quantum state \cite{wootters1982single}. This so-called \textit{optimal cloning} can be achieved with a cloning machine, where one photon is cloned to two imperfect \textit{copies}. Here, the cloning fidelity $F^d_{clon}$, (the overlap of the clones to the original state), is limited to the upper bound of $F^d_{clon}=\frac{1}{2}+\frac{1}{(1+d)}$. This dependence on the dimensionality shows that the fidelity of cloned qubits can be as high as $F^2_{clon}=5/6$, while states with higher dimensionality $d$ can only be cloned with lower fidelities down to 50\% in the limit of $d\rightarrow \infty $. This reduction of the cloning fidelity for qudits depicts nicely the superiority of high-dimensional quantum cryptography and the increased error thresholds as it is more difficult for an eavesdropper to hack a high-dimensional key distribution scheme.  

Hence, additionally to the increased information capacity qudit-states are more robust to background noise and hacking attacks when used in quantum cryptography \cite{bechmann2000quantum,cerf2002security}. To ensure the security of an established quantum link, which can then be used to share a random encryption key, it is essential to stay below a certain error threshold. When this error bound is met, no eavesdropper could have attacked the transmission and security is ensured by quantum physical laws. It is important to note, that it does not matter if an actual eavesdropper had access or the errors were cause by strong noises. This feature of quantum cryptography holds for both entanglement based systems \cite{ekert1991quantum} as well as prepare-and-measure QKD schemes \cite{bennett1984quantum, bennett1992quantum}. For qubit systems, this bound is proven to be as low as approximately 11\% or 12.6\% if two or three mutually unbiased bases (MUBs) are taken into account, respectively \cite{shor2000simple,scarani2009security,bruss1998optimal}. However, if higher-dimensional quantum states are used, this requirement can be weakened, e.g. with ququarts already up to nearly 19\% or 22\% errors can be tolerated (again, for two or three MUBs), respectively \cite{scarani2009security,cerf2002security,lo2001proof,bradler2016finite}. This threshold can be as large as 50\% for an infinite dimensional state, a behavior for which optimal cloning already gave a hint.

\subsection{Quantum Communication without monitoring signal disturbance}
It was discovered recently that high-dimensional systems can also be beneficially for prepare-and-measure quantum cryptography applications if only qubits are encoded and are somewhat \textit{hidden} in the enlarged high-dimensional Hilbert space. The two-dimensional quantum information, which after successful performance of the protocol is used to establish a secure key, is encoded in the relative phases between the utilized high-dimensional states. It was proven that after a clever randomized measurement at the receiver and a large enough Hilbert space, the monitoring of the signal disturbance can be neglected \cite{sasaki2014practical}, in contrast to other prepare-and-measure protocols such as the seminal BB84 \cite{bennett1984quantum}. In simplified terms, this means that renouncing the advantage of a large information capacity per photon and still using complex qudit states facilitates a secure transmission in highly noisy channels.

\subsection{Larger Violation of Local-Realistic Theories and its advantages in Quantum Communication}\label{localRealism}

John Bell has shown in 1964 that the predictions of specific quantum mechanical measurements on qubits cannot be explained in a local realistic manner \cite{bell1964einstein}
\footnote{Local realism is the worldview in which physical properties of objects exist prior to and independent of the measurement, and physical influences are bounded by the speed of light. A violation of Bell's inequality (which have been demonstrated recently in a loophole-free way \cite{hensen2015loophole, giustina2015significant, shalm2015strong, rosenfeld2017event}) shows that such a worldview can no longer be maintained in the face of experimental evidences.}. First generalizations to multi-level systems in the early 1980s by Mermin and Garg \cite{mermin1980quantum, garg1982bell} have shown that the conflict between local realism and quantum mechanics diminishes for larger dimensions. Interestingly, those results were interpreted as a \textit{necessary emergence of local realism in the classical limit}, or a quantum-to-classical transition for large quantum numbers. A decade later, in 1992, Peres was able to show for the first time a violation of local realism that is constant for arbitrary large dimensions \cite{peres1992finite}. It took nearly another decade until Kaszlikowski and colleagues have shown the first high-dimensional violation of local realism that is larger than those in two dimensions \cite{kaszlikowski2000violations}\footnote{A larger violation of Bell type inequalities means that the violation is more robust against noise.}. Their numerical approach has shown that for 3 $\leq$ d $\leq$ 9, Bell-type inequalities can be found where the violation of local realism increases as the dimension of the system increases. Soon afterwards, inequalities for violating local realism for arbitrarily high-dimensional systems have been found by Collins and colleagues (often referred to as CGLMP inequality nowadays) \cite{collins2002bell}. While analytical results are still missing, numerical estimates suggest that the violations increase monotonically with the number of dimensions.

The violation of local realism is not only interesting for fundamental reasons, but has a direct application in quantum communication protocols. These advantages go beyond the more stringent cloning fidelities discussed chapter \ref{chapterSecurityQC}. In entanglement-based QKD protocols, the presence of an eavesdropper is excluded by violating a local realistic inequality. As higher-dimensional systems allow for larger violations, those systems can supersede two-dimensional implementations of QKD. Analogous as before, one can imagine situations where qubits can not be used to generate a secure key, whereas a high-dimensional system still enables to securely distribute an encryption key. We explain this now in two concrete examples.

Huber and Pawlowsky \cite{huber2013weak} investigated an entanglement based quantum key distribution scenario \cite{ekert1991quantum}. There, Alice and Bob measure entangled photon pairs, which they use for creating the key and verifying the security of the transmission. For generating the key, they both measure in perfect correlations (such as H/V, D/A and R/L in polarisation). To verify the security, Bell-measurements need to be performed (such as H+22.5$^o$/V, H-22.5$^o$/V, D+22.5$^o$/A, D-22.5$^o$/A). If the eavesdropper knows when the Bell-measurement is performed, she would not look at those photons (to not disturb them), and only look at the photons for the generation of the secret key. Therefore, the choice when the Bell measurement is performed has to be random. If the randomness is weak, the eavesdropper could still try to hide her appearance \cite{bouda2012weak}. Above a certain threshold, weak randomness prevents therefore a secure quantum key distribution. Interestingly, because the violation of high-dimensional Bell inequalities can be much stronger than two-dimensional violations, the threshold of acceptable loss of randomness is significantly larger for high-dimensional systems. This leads to the situation where two-dimensional quantum key distribution (for instance, performed with polarisation of photons) is hopeless while high-dimensional systems would still permit a secure channel \cite{huber2013weak}.

Device-independent quantum key distribution (DI-QKD) requires the detection of a large fraction of the distributed entangled states. The smallest known fraction  for closing the so-called \textit{detection loophole} is 66.7\%, based on non-maximally entangled states as shown by Eberhard \cite{eberhard1993background}. Reaching this bound is quite challenging for photons, and has only been exceeded in 2013 \cite{giustina2013bell, christensen2013detection}. However, it was shown by Vertesi, Pirono and Brunner that for d=4, the detection efficiency required for overcoming the detection loophole can drop by almost 5\% to 61.8\% \cite{massar2002nonlocality, vertesi2010closing}. Surprisingly, results for larger dimensions are not known so far.

\subsection{Quantum Computation with QuDits}

Using qudits can also have advantages in quantum computing. For example it has been shown that distillation of important resource states for quantum computations (so-called Magic states) can be done in a way that have no direct analogue for qubits and can be several orders of magnitude more efficient \cite{campbell2012magic}. In another approach Bocharov and colleagues investigate the Shor algorithm with qutrits, and find the promising result that qubits encoded in the slightly larger Hilbert space of a qutrit allow for a more efficient circuit structure than using natural qutrit gates \cite{bocharov2016factoring}.

\section{Recent Developments in High Dimensional Quantum Information with OAM}
After having discussed a few advantages that qudit systems offer, we will now turn to recent developments of the broad and vivid activity of high-dimensional quantum information using the OAM of photons. 

\subsection{Creation of High Dimensional Entanglement}
Photon pairs entangled in their orbital angular momentum can be produced using a spontaneous parametric down-conversion process in a nonlinear crystal \cite{mair2001entanglement}. Due to conservation of OAM (when the pump photon has $\ell_p=0$), the two photons have opposite OAM $\ell_s = - \ell_i$ such that the resulting two-photon state can be written as 

\begin{align}
\ket{\psi}&=a_0\ket{0}_s\ket{0}_i+a_1\ket{1}_s\ket{-1}_i+a_{-1}\ket{-1}_s\ket{1}_i + \cdots \nonumber \\
&=\sum_\ell a_{\ell}\ket{\ell}_s \ket{-\ell}_i
\label{eq:SPDCstate}
\end{align}
where $a_{\ell}$ are coefficients, which depend on the focusing parameters and crystal dimensions. This process has been employed in the first quantum entanglement experiment with OAM \cite{mair2001entanglement} and has been the workhorse for nearly all quantum experiments using OAM entanglement as one gets high-dimensional entanglement nearly \textit{for free}.

A stringent limitation however is that there are only very few methods to tune the parameters $a_{\ell}$. These parameters decrease significantly for higher-order modes, as has been investigated in detail by Miatto and colleagues in \cite{miatto2011full}. Experimentally, it has been shown that by chosing the focussing parameters in an optimal way, one can increase the higher-order modes \cite{romero2012increasing}, however not to a flat distribution. As an effect, the resulting entangled state is not maximally entangled which limits its usefulness in quantum experiments.

An interesting recent advancement has been introduced by Zhang and colleagues \cite{zhang2016engineering}. They show how the usage of two-photon interference (Hong-Ou-Mandel effect \cite{hong1987measurement}) can be used to generate a large set of high-dimensional states by quantum state filtering. 

In another experiment, high-dimensional entanglement was generated at first in the path degree of freedom, which is simpler to custom tailor \cite{schaeff2015experimental}, and afterwards transferred to OAM states \cite{fickler2014interface}. Because path encoded states are compatible with integrated optics, this technique might also serve as an interface between waveguide structures and OAM modes of photons.

\begin{figure}[t]
\includegraphics[width=0.5 \textwidth]{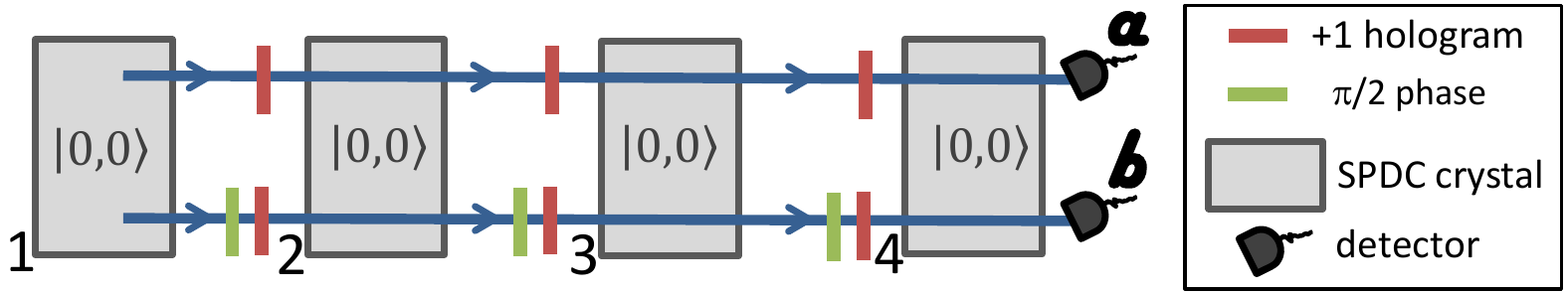}
\caption{An experimental scheme for creating a 4-dimensionally entangled two-photon state, based on entanglement by path identity \cite{krenn2017entanglement} (image from there). Four crystals are pumped coherently and the paths of the down-converted photons are made identical. If the experimental conditions are such that there is in principle no information about the photon pairs origin (i.e. in which of the four crystals it has been created), then the pair is in a coherent superposition of being created in either of them. One can then manipulate the mode number and phase of the individual terms in the state. In this example, a hologram between each crystal shifts the mode by one and a phase of $\pi/2$ is added such that the resulting state is a 4-dimensional Bell state $\ket{\psi}=1/2\left(\ket{0,0}+i\ket{1,1}-\ket{2,2}-i\ket{3,3}\right)$.}  
\label{FigPathIdentity}
\end{figure}

Another alternative approach that allows for the generation of arbitrarily, high-dimensional two-photon states has been proposed recently \cite{krenn2017entanglement}. The idea is, that the creation of a pair of photons happens in a coherent superposition of several crystals. Between the crystal, the mode number, phase and amplitude can be adjusted arbitrarily. The generation of an arbitrary 4-dimensional state is shown in Fig. \ref{FigPathIdentity} as an example. In general $d$ crystals are pumped coherently such that one pair of photons is created in each crystal. As it is not known in which of the $d$ crystals it is created, the state is in a coherent superposition of being created in either of them. Manipulation of the mode number and phase between the crystals allow for creating arbitrary two-photon entangled states --  e.g. a $d$-dimensional Bell state. As only the Gaussian mode of the down-converted photons are used, the count rates of this method can be significantly larger than for other schemes. The same method can also be used to create very general high-dimensional multipartite entangled states \cite{krenn2017entanglement} with interesting connections to Graph Theory \cite{krenn2017quantum}. Interestingly, the method has been discovered using a computer algorithm which automatically designs new quantum experiments \cite{krenn2016automated}. 

\begin{figure*}[htbp]
	\centering
	\includegraphics[width=0.95\textwidth]{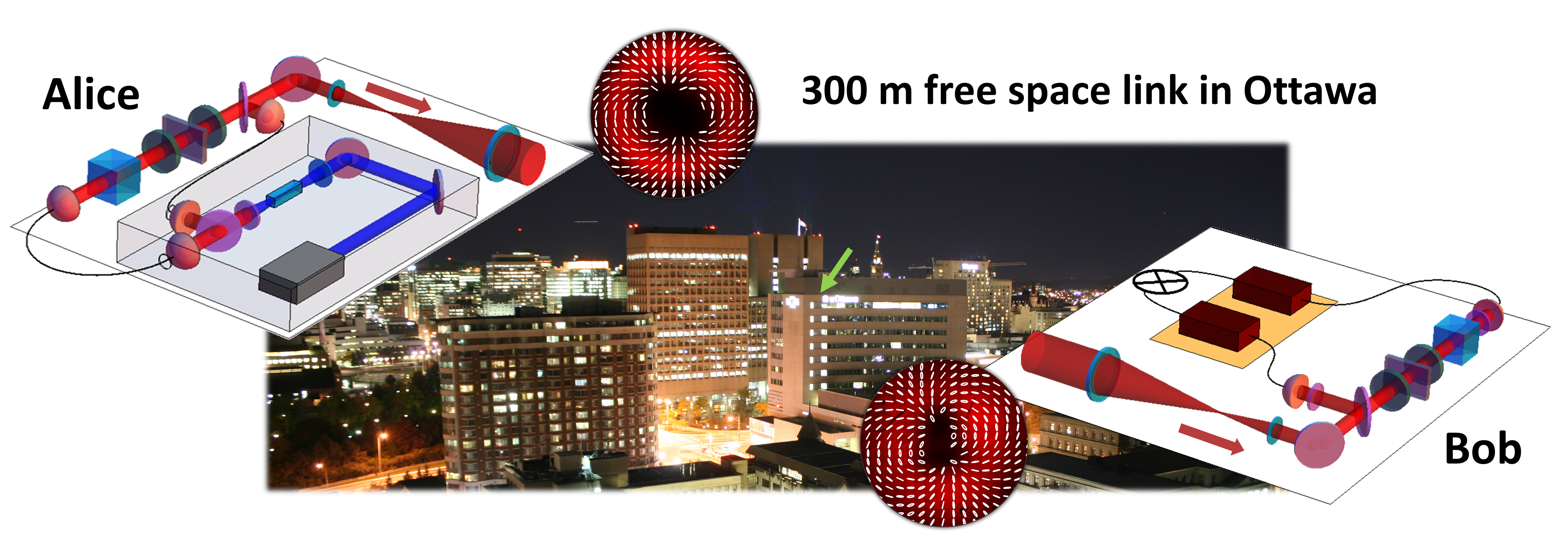}
	\caption{High-dimensional QKD over a 300 m long intra-city link in Ottawa \cite{sit2016high}. Alice prepares non-degenerate pairs of photons in a down conversion process and splits them depending on the wavelength (grey box). Ququart states imprinted on the signal photon are prepared by means of an appropriate set of a polarizing beam splitter (cube), wave plates (green) and a q-plate (violet). A recording of an example state with a complex polarization pattern is shown in the upper inset. The two photons are then recombined and sent to Bob via the moderate turbulent link over 300 m. Bob splits the two photons by their wavelength, measures the received state of the signal photon with same set of devices (received pattern shown in the lower inset) and heralds its arrival by detecting the idler photon. An uncorrected error rate of 14\% was found corresponding to 0.39 bits per photon and which can only considered to be secure for high-dimensional states as the threshold for qubits is 11\% (for ququarts the bound is 18\%). The background photo shows a night view of Alice's building seen from Bob's receiving station, the green arrow points to the actual location of the sender.}
	\label{fig:freespace}
\end{figure*}

\subsection{Unitary Transformations}\label{sectionTrafo}
While the generation of entangled photons with OAM is well-developed and in the focus of different research efforts, the capability of transforming modes, which is essential for quantum informational tasks, is still in its infancy.
In contrast to path-encoding of photonic qudits, where re-programmable chips exist that can perform arbitrary unitary operations chips \cite{schaeff2015experimental,carolan2015universal}, for OAM only a limited set of transformations are known.

One proposed idea towards achieving arbitrary transformations from 2010 is using a succession of reflections from a programmable deformable mirror device \cite{morizur2010programmable}. By using a feedback signal and a stochastic optimization, it was shown that already after three reflection a good transformation between gaussian mode and a third order Hermite-Gauss mode can be achieved. It will be interesting to see whether this approach can be used to perform mode-dependent transformations with precise control over phases at the quantum level.

A more recent technique might be another promising candidate to realize general transformations on OAM states. It has been demonstrated that using controlled random scattering, i.e. properly shaping the wavefront before a strong scattering process, sorting arbitrary spatial mode structures (including radial $p$ modes) can be achieved in a coherent way \cite{fickler2017custom}. Improvements on the overall loss of the process as well as possible limitation of mode transformations will be an important next step in order to successfully apply this technique in quantum experiments.    

Another route to a multiport device for OAM modes is to build the required unitary transformation out of basic components. Here, the unitary operation that permutes all utilized OAM states in a cyclic manner, i.e. a cyclic ladder transformation or high-dimensional X-gate, serves as part of the building blocks for a general device. Together with the high-dimensional Z-gate, which introduces mode dependent phases, any unitary transformation can be constructed \cite{asadian2016heisenberg}. While the Z-gate for OAM modes directly corresponds to a rotation and thus can be realized by a single optical device \cite{leach2002measuring,de2005implementing}, the general experimental implementation for a cyclic transformation is not known yet. However, for dimension d=4 and d=5 (and d=3, d=6 and d=8 when one uses polarisation), an all-optical linear setup has been found \cite{krenn2016automated,chen2017realization} and implemented recently \cite{schlederer2016cyclic,babazadeh2017high}. A combination of several of these basic gates are required for arbitrary transformations, thus integrating them in a stable manner will be required. Furthermore, X-gates in other dimensions than those found will be required to exploit the full capability of the high-dimensional space.

\subsection{Optimal Quantum Cloning}
The experimental procedure for cloning of qubits was established in different experiments more than a decade ago \cite{scarani2005quantum}, no high-dimensional state has been optimally cloned until recently \cite{nagali2010experimental,bouchard2017high}. The experiments use two-photon Hong-Ou-Mandel interference \cite{hong1987measurement} of OAM modes at a beam splitter \cite{nagali2009optimal} to achieve optimal cloning of high-dimensional OAM states up to the dimension 7 \cite{bouchard2017high}. It was not only shown that it is possible to clone any qudit state from any mutually unbiased bases, hence the cloning machine is universal, but it was also demonstrated that the measured cloning fidelities follow the formula described in chapter \ref{chapterSecurityQC} and hacking a high-dimensional QKD scheme with a cloning attack does result ins larger errors, thus can be easier detected. 

\subsection{Long Distance High Dimensional QKD}
A promising application of high-dimensional OAM states is their use in quantum cryptography schemes. While many proof-of-principle experiments have been conducted successfully in laboratories around the world \cite{groblacher2006experimental, mafu2013higher, mirhosseini2013efficient, vallone2014free}, current research focuses on bringing the technology closer to real world applications by enlarging distances over which secure keys can be distributed. Specially developed vortex fibers \cite{bozinovic2013terabit} might enable fiber networks that rely on OAM encoding, however, the implementation in quantum experiments has not been demonstrated yet. Moreover, expanding such networks to global distances might be facilitated via free space links. Here, qudits help to increase data rates as well as lower the requirements for correction of induced errors, both being strongly required in outdoor environments due to inherent loss and background noise.  After recent successful tests of classical communication schemes using spatially structured light over distances up to 143 km \cite{gibson2004free,wang2012terabit,krenn2014communication,ren2016experimental,krenn2016twisted}, first free-space quantum experiments using spatial structures have been demonstrated. In the first of these experiments, photons entangled in their OAM degree of freedom have been sent over an intra-city free-space link over 3 km in Vienna \cite{krenn2015twisted}. While it was possible to demonstrate the survival of OAM entanglement for photons up to $\ell=|2|$, the quantum states were also only two-dimensional. Recently, the first high-dimensional quantum states have been used to perform an free-space intra-city QKD experiment over 300 m in Ottawa (see Fig. \ref{fig:freespace}) \cite{sit2016high}. 4-dimensional quantum states, i.e. ququarts, were imprinted on heralded single photons using a combination of OAM and polarization. The photons were then transmitted from Alice to Bob following the standard high-dimensional BB84-protocol \cite{bennett1984quantum}. Although moderate turbulence was observed, a coupling to single mode fibers, which is required to measure the received state, was achieved with up to 20\% efficiency. More importantly, an uncorrected error rate of around 14\% was found, corresponding to 0.39 bits per sifted photon and well below the threshold of 18\% for ququarts. The additional transmission of the heralding photon from Alice to Bob and its usage to gauge time periods of weak turbulences has further improved the extracted key rate to 0.65 bits per sifted photon. On one hand, the results show that free-space QKD with spatially structured photons is feasible and high-dimensional states can be superior over qubit encoding. On the other hand, they also indicate that the extension to larger distances as well as higher data rates will require the implementation of adaptive optics to compensate for the detrimental effects of atmospheric turbulences \cite{ren2014adaptive}. Hence, the next steps will be to develop appropriate compensation schemes, to test quantum states of higher dimensions, e.g. with pure OAM encoding, and distribute high-dimensionally entangled photons for large quantum networks.

\subsection{Quantum Walk}
Another exciting example that shows the advantages of OAM implementations over other degrees-of-freedom, is the implementation of a quantum (random) walk (QW). In a classical random walk on a 1D chain, a walker randomly decides (for instance by flipping a coin) at every step to move left of right. The QW is obtained when both the coin and the walker are quantum objects. Despite the seemingly simple setting, QW experiments have a great potential. It has been demonstrated that QW allows for dynamic simulation of complex quantum systems, for the implementation of quantum search algorithms \cite{shenvi2003quantum} and even universal quantum computation \cite{childs2009universal}.

In a photonic system, a quantum walk is often considered as a photon, which enters a beam splitter network with its polarisation acting as the coin operation \cite{broome2010discrete,schreiber2010photons}. As a QW needs to be interferometrically stable (i.e. the paths must be stable much below the wavelength of the photons), larger networks of interferometers are usually integrated into photonic waveguides \cite{peruzzo2010quantum, sansoni2012two}.

\begin{figure}
\includegraphics[width=0.5 \textwidth]{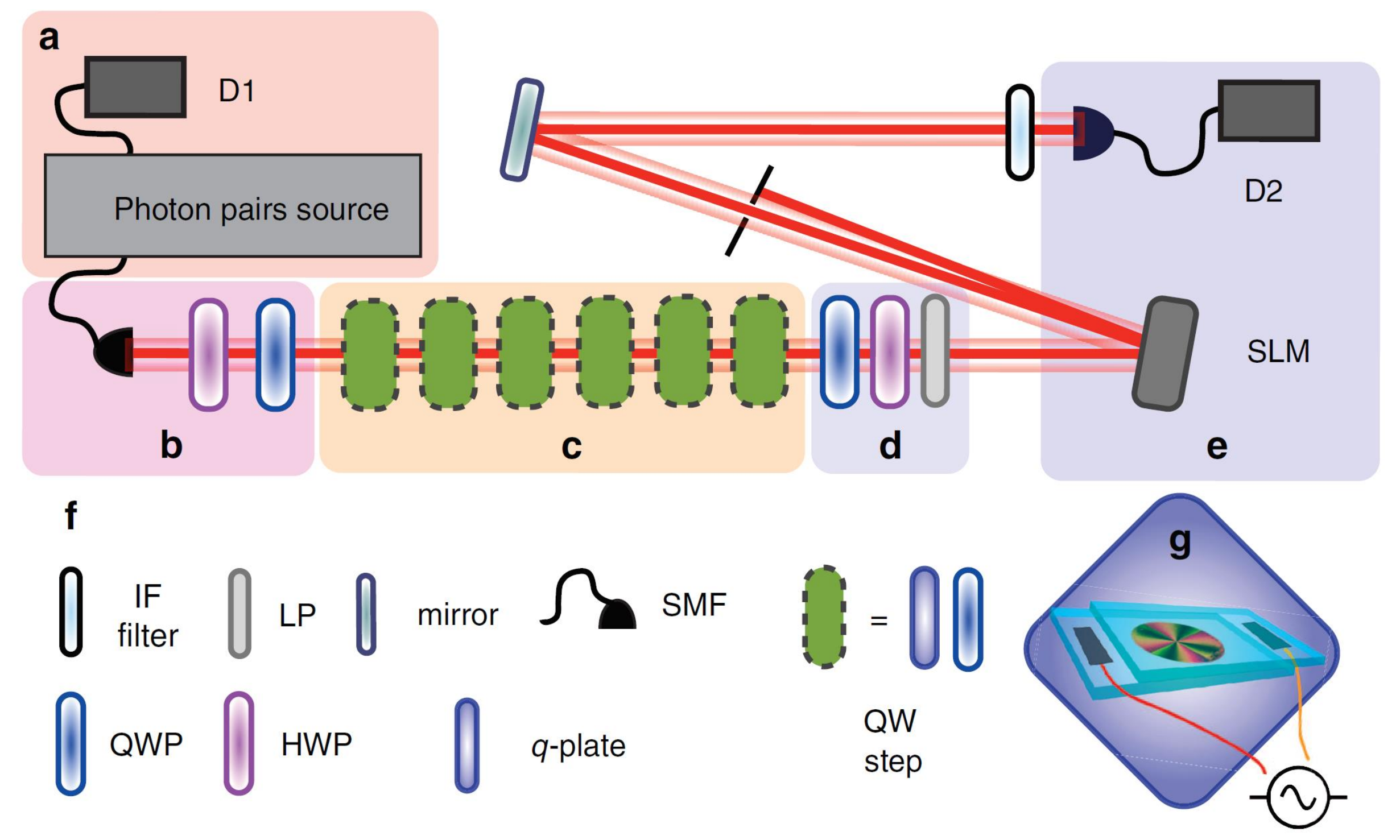}
\caption{Sketch of a quantum walk in OAM with six steps, used in \cite{cardano2016statistical} (image from there). \textbf{a} A source of two indistinguishable photons. \textbf{b} and \textbf{d} are used to control the polarisation of the light before and after the QW. \textbf{c} shows the six steps of the quantum walk and \textbf{e} shows the detection system based on SLMs and single-mode fiber.}  
\label{FigRandomwalk}
\end{figure}

In 2010, Zhang and colleagues showed that one could use OAM (instead of path) and polarisation to perform QWs using so-called q-plates \cite{zhang2010implementation}, which are devices developed by Marrucci and colleagues \cite{marrucci2006optical} that can couple polarisation and OAM of light. Such schemes have the great advantage that integrating the walker into photonic waveguide circuits is not necessary anymore, as the phase between different OAM modes does not fluctuate during propagation (see Fig.\ref{FigRandomwalk} for an sketch of such a setup). In 2015, Cardano and colleagues have used a set of five q-plates to perform a quantum walk with 5 steps \cite{cardano2015quantum} and investigated the dynamics of the propagation for both single and two-photon inputs. Other interesting applications for QW are the exploration of topological phases in 1-dimensional and 2-dimensional systems \cite{kitagawa2010exploring}. Those generalized geometric phases introduced by Berry \cite{berry1984quantal} play a key role in solid-state physics, such as the description of the quantum Hall effect \cite{thouless1982quantized} (see also \cite{lu2016topological}). Cardano and colleagues were also able to use their multistep q-plate system for observing a topological phase transition \cite{cardano2016statistical} and topological invariants \cite{cardano2017detection} in a one-dimensional system.

\subsection{Quantum Teleportation of Multiple-Degrees of Freedom of a Single Photon}\label{QTeleportExperiment}

\begin{figure}[ht]
	\centering
	\includegraphics[width=0.45\textwidth]{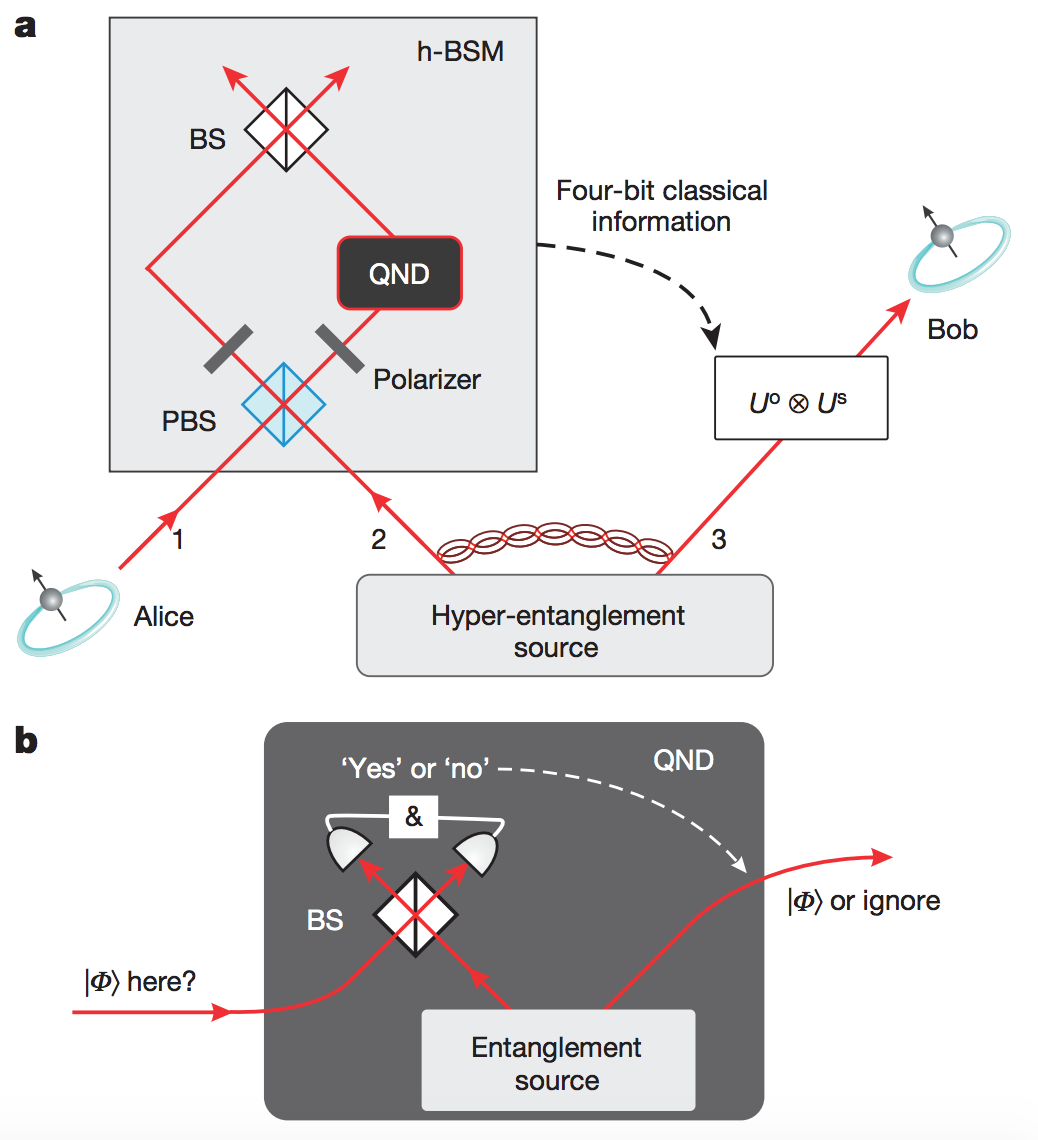}
	\caption{\textbf{a)} Experimental scheme to simultaneously teleport two degrees of freedom of a single photon, as described by \cite{wang2015quantum} (image from there). A hyper-entanglement source produces photon pairs entangled in their polarisation and OAM degree-of-freedom. The polarisation degree-of-freedom is teleported at the polarising beam splitter (PBS). The two polarizers then erase the polarisation information on the photons. To ensure that the Bell state measurement for the polarisation teleportation succeeded, a quantum-non-demolition measurement (QND) is performed. This QND itself is again a teleportation scheme, see inset \textbf{b)}. Then the OAM degree-of-freedom is teleported with a Bell-state measurement performed at the upper beam splitter. The results of both Bell-state measurements are then encoded in four classical bits and send to Bob who can perform an according unitary transformation to restore the initial quantum state from Alice.}
	\label{fig:teleportation}
\end{figure}

Quantum teleportation, proposed in 1993 by Bennet and colleagues \cite{bennett1993teleporting}, is an important quantum protocol with applications in quantum network and quantum computation. It allows Alice and Bob (if they share an entangled state) to distribute an arbitrary quantum state without sending it, just by sending classical information.

The concept is shown in figure \ref{fig:teleportation}b, for a two-dimensional system. There, Alice and Bob share a maximally entangled Bell state. Alice has an arbitrary quantum state $\ket{\phi}$ which she wants to transmit to Bob. She performs a Bell-state measurement between her entangled photon and $\ket{\phi}$. Then she announces the result via a classical channel to Bob. If her result was a $\ket{\phi^+}$ state, the initial state $\ket{\phi}$ has been transmitted to Bob successfully. If the outcome of the state measurement is different, Bob can perform a transformation to get the original $\ket{\phi}$ (or he ignores the photon which reduces the efficiency of the implementation). The first experiment implementing quantum teleportation was performed 1999 by Bouwmeester and colleagues ~\cite{bouwmeester1997experimental} utilizing the polarisation of photons. Since then, quantum teleportation has been performed in various different physical systems with ever improving quality -- very recently even between a satellite and a ground station in China \cite{ren2017ground}. All of these experiments have in common that they only teleport one degree-of-freedom of the quantum particle, and only in a two-dimensional system -- thus only a small part of the full information of the quantum system was transferred.

A recent experiment by Wang and colleagues \cite{wang2015quantum} has overcome this limitation for the first time, by showing the simultaneous teleportation of two degrees of freedom of a single photon. The basic idea behind the teleportation of multiple degrees of freedom is to successively teleport one degree after the other. In experiments so far, the incoming quantum state was destroyed during the teleportation, thus no second teleportation was possible. The clever idea to get around this is to use a quantum-non-demolition measurement (QND). In this QND measurement, the quantum state is not destroyed and thus the remaining quantum information, which is stored in the other degrees of freedom, is still intact and can be teleported in the next steps. Interestingly, this QND itself is another teleportation protocol. Thus, the authors used teleportation itself as QND to enable the teleportation of a second degree-of-freedom of the particle. The measured fidelities $F=\text{Tr}(\rho|\psi\rangle\langle\psi|)$ between the experimentally teleported state $\rho$ and the theoretically expected state $|\psi\rangle$ range from $0.68\pm0.04$ to $0.62\pm0.04$. The classical limit for optimal quantum state estimation of a pure state~\cite{hayashi2005reexamination} is given by $F_\text{opt}^\text{cl}(d=4)=\frac{2}{1+d}=0.4$, where $d$ is the dimension of the system. Thus, the results show the successful implementation of a quantum teleportation scheme for a polarisation-OAM composite state of a single photon.

In the future, it will be important to teleport more levels of a single high-dimensional degree-of-freedom, and to teleport more than two degrees-of-freedom, in order to transmit more information stored in a quantum state.

\subsection{Experimental Creation of a Greenberger-Horne-Zeilinger State in Three-Dimensions}

Three particle GHZ states in two dimensions have been first discovered and investigated in 1989 by Greenberger-Horne-Zeilinger~\cite{greenberger1989going,greenberger1990bell} and later generalized to N particles by Mermin~\cite{mermin1990extreme}. Since their discovery, GHZ states play an essential role in quantum information, especially in the classification of entanglement, communication-complexity problems, quantum-error-correction schemes for quantum computing and also for novel experimental tests of local-realistic theories. 

GHZ states have been intensively studied experimentally as well as theoretically. All experimental studies focused thereby on increasing the number of involved particles rather than increasing the dimensionality of the single particles involved. Recently, a first step towards achieving this goal was reported by entangling three particles out of which two were entangled in three dimensions, while the third particle was encoded as a qubit \cite{malik2016multi}. This asymmetric arrangement already shows that multi-partite, high-dimensional quantum states offer novel features of entanglement \cite{huber2013structure}. Finally -- in a very recent experiment -- the first completly high-dimensional multipartite state was created. It is a generalisation of GHZ state which goes beyond qubits \cite{3d-ghz}.

\begin{figure}[ht]
	\centering
	\includegraphics[width=0.47\textwidth]{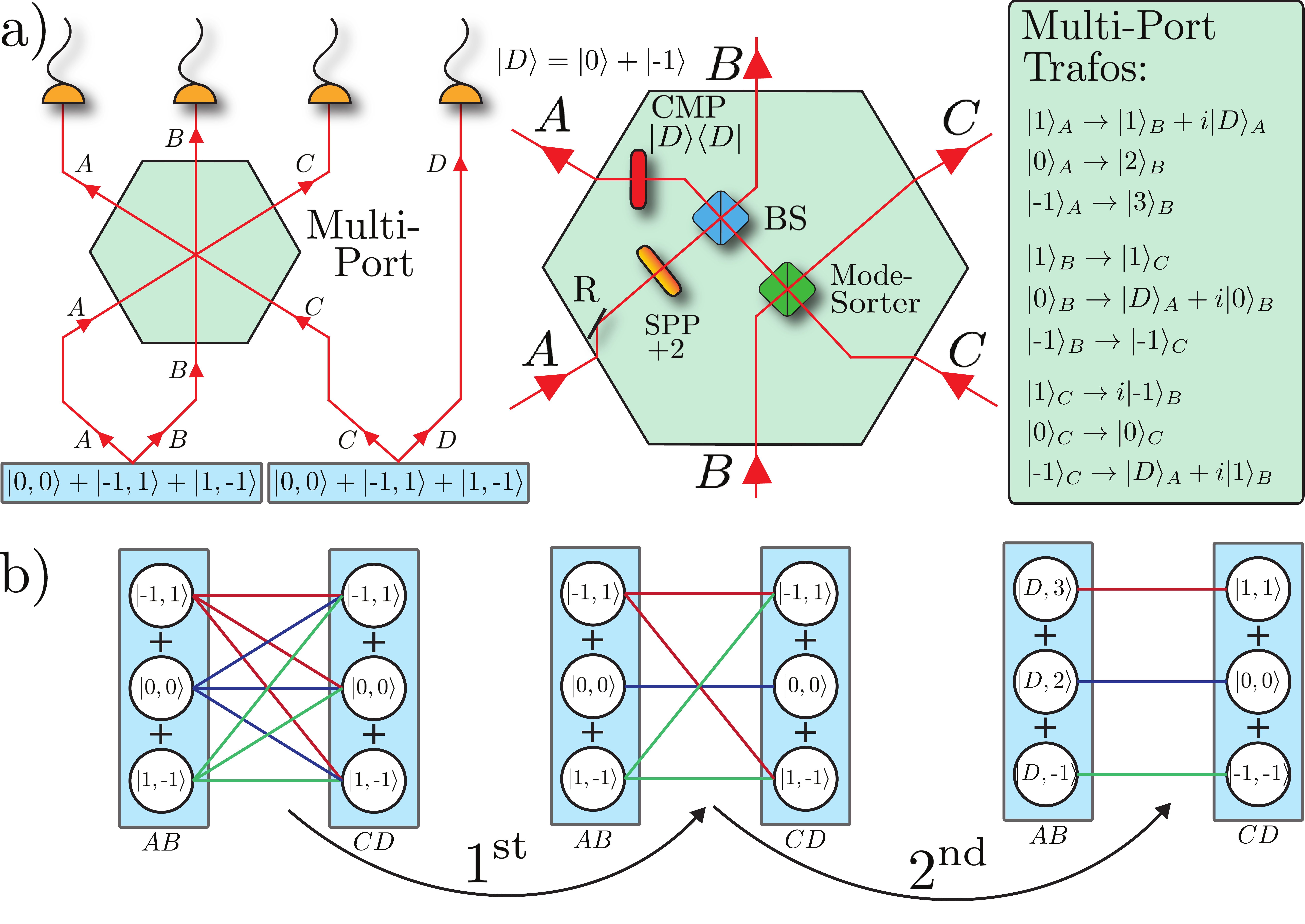}
	\caption{\textbf{a)} Experimental scheme to generate a three-photon and three dimensional GHZ state \cite{3d-ghz} (image from there). Two non-linear crystals each produce a pair of photons that are three dimensionally entangled. Three out of the four photons (A,B,C) enter a multiport which transforms each photon according to the transformations depicted in the figure. Interestingly, for each input port (A,B,C) of the multiport a different OAM mode $\ket{-1},\ket{0},\ket{1}$ exits in a superposition of two different paths. The multiport consists of a reflections (R), spiral-phase-plate (SPP) that add a constant value of $\ell=2$, a mode sorter that sorts OAM modes according to their parity (even/odd), a 50/50 beam splitter (BS) and a coherent mode projection (CMP). \textbf{b)} Graphical representation of the three dimensional GHZ state generation. If the two non-linear crystals emit a photon pair coherently at the same time, then there exist $3\times 3=9$ probability amplitudes in total. These nine probability amplitudes are depicted by the red, blue and green connection lines. In the $1^\text{st}$ step, the mode sorter eliminates all probability amplitudes between even and odd terms conditioned on the simultaneous detection of a photon in each detector. In the $2^\text{nd}$ step, a Hong-Ou-Mandel two photon interference at the BS eliminates the green connection line between $\ket{1,-1}_\text{AB}$ and $\ket{-1,1}_\text{CD}$. Finally, the last remaining unwanted connection line is eliminated by rerouting the incoming photons such that no simultaneous detection event in all detectors can occur. Photon A always exits the multiport in the $\ket{D}$ state and is therefore not entangled with the other three anymore. This leaves photons B,C and D in a three dimensional GHZ state.} 
	\label{fig:3d-ghz}
\end{figure}

While for two-dimensional GHZ states a clear recipe to increase the number of particles exists, this does not hold for increasing the dimensionality. The authors utilized the computer algorithm \melvin \cite{krenn2016automated} to find an experimental setup to create a three-dimensional GHZ state $\ket{\psi}=\left(\ket{0,0,0}+\ket{1,1,1}+\ket{2,2,2}\right)/\sqrt{3}$. Despite the simple correlations present in this state the actual experimental setup, depicted in Fig.~\ref{fig:3d-ghz}, is neither intuitive nor simple.
The main idea is to use two sources of three-dimensional entangled photon pairs and combine them in such a way that one trigger photon heralds the 3-dimensional GHZ state in the other three photons. If the two photon pairs are emitted at the same time, the overall four photon state shows correlations between each of the 3x3=9 terms of the quantum state. These correlations are depicted in Fig.~\ref{fig:3d-ghz}~b) as red, green and blue lines. At first, there are nine connecting lines, while in the final three dimensional GHZ only three connections exist. Thus six of the nine probability amplitudes need to be coherently suppressed in the experimental setup. This can be achieved with a three input and three output multiport. The detailed transformations of the multiport are given in Fig.~\ref{fig:3d-ghz}~a). The basic idea behind the multiport is to utilize an even/odd OAM mode sorter \cite{leach2002measuring}, a Hong-Ou-Mandel effect, a coherent mode projection and spiral phase plates (SPP) to suppress all but three possible correlations. Finally, a simultaneous detection of four photons results in a three-dimensional GHZ type entangled state for photons B,C and D. The experimental fidelity ($F_\text{exp}=\text{Tr}(\rho|\psi\rangle\langle\psi|)$) of $F_\text{exp}=75.2\%\pm2.9\%$ shows genuine three-photon and three-dimensional entanglement.

Experiments like the ones described above will allow for the investigation of the strongest form of contradictions to local realism -- namely deterministic violations of local realism as introduced by Greenberger, Horne and Zeilinger (GHZ) \cite{greenberger1989going}. Generalizing GHZ-violations to higher dimensions have been surprisingly difficult and has only been achieved very recently \cite{ryu2013greenberger, ryu2014multisetting, lawrence2014rotational, lawrence2017mermin}. Those constructions departure significantly from their two-dimensional counterparts, and it seems as if they would help revealing the real nature of deterministic violations of local realism.

\section{Future Challenges}
Finally, we will discuss a few open questions and challenges that will further advance the field of high-dimensional quantum states encoded through the OAM of photons.

\subsection{Long-distance distributing high-dimensional OAM states}
The undisturbed distribution of high-dimensional quantum states is one of the most important tasks, if applications should advance beyond laboratory proof-of-principle demonstrations. Here, two different approaches are presented, fiber and free-space transmission, each face their own individual challenges.

A first step towards fiber-based long-distance distribution of OAM states, has been achieved in developing a vortex fiber that transmits OAM states of light \cite{bozinovic2013terabit}. Although this fiber has been around for years, the implementation in a quantum optical setup is still yet to be demonstrated. Moreover, high-dimensional transmission is not possible, but two other approaches are conceivable: One approach would be carefully designing OAM multimode fibers, in which the modes are degenerate but do not couple to each other \cite{gregg2015conservation}. Another possibility is the fabrication of standard multimode fibers, in which intermodal coupling has to be pre- or post-compensated or reduced by an appropriate choice of the modal set \cite{huang2015mode,carpenter2015observation} (recently, up to 50 kilometers \cite{wang2016characterization}). As each will have its own benefits and drawbacks it will interesting to be pursuit tested in laboratories as well as real-world scenarios. 

As described earlier, the second method of distributing OAM encoded qudits, namely free-space distribution, is already a step ahead. However, it is not a coincidence that the only qudit experiment was performed over a short distance (ququarts over a distance of 300 m \cite{sit2016high}). While atmospheric turbulences might not completely destroy the transmitted state, even after 143 km \cite{krenn2016twisted}, it introduces considerable crosstalk between different modes, mainly due to beam wander and distortions. Therefore, adaptive optics for the correction of wave front distortions is indispensable and first steps are currently developed \cite{ren2014adaptive}. A successful demonstration would allow expanding the dimensionality of the state along with enlarging the distances over which they are distributed, maybe even between earth based ground stations and satellites \cite{yin2017satellite,ren2017ground,liao2017satellite}.

Progress in these questions could lead to global quantum networks where quantum information is shared via high-dimensional quantum states encoded OAM.

\subsection{Quantum communication with photons in other modes}
So far, we only discussed the OAM of photons. However, Laguerre-Gauss modes have also a second quantum number -- the radial quantum number \cite{karimi2014radial,plick2015physical} --, which can be used in quantum experiments \cite{karimi2014exploring,krenn2014generation}. It would be interesting to develop techniques to manipulate, detect and investigate properties of the radial modes to enable their use in quantum communication experiments. The investigation of OAM modes in comparison with other modes would be interesting. Do other mode families have different advantages? Self-healing properties of Bessel modes have been shown impressive results in the laboratory \cite{mclaren2014self}. Can this apparent advantage also be exploited in real outdoor experiments? Do other, more complex beams have advantages in real-world quantum communication experiments -- for example Ince-Gauss modes \cite{krenn2013entangled} which have only single-charged vortices in their phase profile, or Airy beams \cite{siviloglou2007observation} which tend to freely accelerate during propagation?

\subsection{Arbitrary transformations of OAM modes and two-qudit quantum gates}
In chapter \ref{sectionTrafo}, three different approaches have been demonstrated which have the common goal of performing unitary transformations. While progress has been made in the last few years, it is still unclear how to perform arbitrary transformations with high fidelity, near unitary efficiency in a fast and reliable way. Such unitary transformations are required for computational tasks and their experimental demonstration might also trigger further investigations into high-dimensional computation and simulation.  
Similarly important for quantum computation tasks are gate operations for single and also for multi-particle systems. While single-qubit transformations and two-qubit gates have been studied, very little is known about high-dimensional two-qudit gates. Such gates are essential to perform the first proof-of-principle investigations of high-dimensional quantum algorithms \cite{campbell2012magic, bocharov2016factoring}.

\subsection{Quantum Teleportation and Entanglement Swapping of High-Dimensional Quantum States}

Another open challenge is to experimentally demonstrate high-dimensional quantum teleportation and high-dimensional entanglement swapping in OAM. The experiment in chapter \ref{QTeleportExperiment} shows the simultaneous teleportation of two degrees of freedom of a single photon, but every degree of freedom is still two-dimensional. This problem is related to the question how to experimentally perform a higher dimensional Bell state measurement. A solution to this problem would not only allow to perform teleportation of high-dimensional quantum states encoded only in the OAM degree of freedom, but also enable other quantum protocols such as entanglement swapping, which is (together with reliable quantum memories) the key ingredient for quantum nodes in quantum networks \cite{duan2001long}.

\subsection{Stronger Violations of Quantum Mechanical vs. Local Realistic Theories}

Finally, we would like to point to an open theoretical question with significant conceptual and experimental implications. Is it possible to find Bell- or Mermin-like inequalities that are more robust against noise or detection inefficiencies in higher dimensions? The known advantages for the two-photon cases have been discussed in chapter \ref{localRealism} -- many questions remain open. Moreover, for high-dimensional multi-photon states, even less is known. The only known violations of local realism are weaker than those for two-dimensional systems \cite{ryu2013greenberger, ryu2014multisetting, lawrence2014rotational, lawrence2017mermin}. In systems where both the number of photons as well as the number of dimensions is larger than two, asymmetric types of entanglement can exist \cite{huber2013structure,huber2013entropy,goyeneche2016multipartite}. There, the connection to violation of local realism and its applications in novel quantum protocols has not been investigated until now.

\section{Final remark}

Within only a few years we have seen a tremendous progress in the capabilities of generation, control and application of high-dimensional quantum states with twisted photons. Among them, we have seen the first outdoor experiments distributing entanglement and high-dimensional quantum keys, we witnessed the first generation of multi-photon entangled states and applications in quantum teleportation. These advances arose from technological progress, but also from a close interaction between theory and experiment. For example the demonstration of novel advantages of high-dimensional systems in quantum cryptography or quantum computation are essential motivations for experiments; new theoretical classifications of multipartite high-dimensional entangled states hint to novel quantum protocols, and computer-designed experiments help realizing complex quantum states and protocols in the laboratories. We believe that the synergy between theorists and experimentalists will remain a key factor in the next few years.

To sum this up -- in the spirit of Richard Feynman, \textit{What I cannot create, I do not understand} -- we believe it is highly desirable to continue investigating high-dimensionally encoded quantum systems -- both on paper and in laboratories.

\bibliographystyle{unsrt}
\bibliography{refs}

\end{document}